\documentstyle[aps,epsfig,multicol]{revtex}
\draft
\input epsf
\begin{document}
\title{\bf Power law load dependence of atomic friction}
\author{C. Fusco\thanks{Author to whom correspondence should be
addressed. Electronic address: fusco@sci.kun.nl.} and A. Fasolino}
\address{Theoretical Physics, NSRIM, University of 
Nijmegen,\\ 
Toernooiveld 1, 6525 ED Nijmegen, The Netherlands
}

\date{\today}
\maketitle

\begin{abstract}
We present a theoretical study of the dynamics of a tip scanning a graphite 
surface as a function of the applied load. 
From the analysis of the lateral forces, we extract the 
friction force and the corrugation of the effective tip-surface interaction 
potential.
We find both the friction force and potential amplitude to have a power law 
dependence on applied load with exponent $\sim 1.6$. We interpret these 
results as characteristic of sharp undeformable tips in contrast to the case 
of macroscopic and elastic microscopic contacts.

\end{abstract}

\begin{multicols}{2}

It is well known that macroscopic 
friction is proportional to the applied load but the load dependence of 
atomistic friction is still under investigation and has been discussed in a 
limited number of experimental and theoretical 
works~\cite{fujisawa1,fujisawa2,ishikawa,sasaki1,zhong,sorensen,schwarz,enachescu}.
Usually the load dependence is described by means of contact mechanics
continuum theories~\cite{schwarz,enachescu} which do not give information 
about the atomic interactions in the sliding contact region and assume a
spherical tip. Moreover, it has been argued that, depending on 
the shape of the tip, power laws with different exponents can be 
found~\cite{schwarz}. 
In this Letter, we present a detailed study of the load dependence of 
atomic-scale friction in the case of a sharp tip-surface contact, finding a 
power law dependence with exponent larger than one. We also show how
the effective tip-surface potential energy barriers can be derived from force 
measurements, thus establishing a connection between friction and the 
interatomic potential.
Recent works using noncontact mode Atomic Force Microscopy (AFM) have shown 
the possibility to reconstruct the tip-surface potential~\cite{holscher1}, 
but they lack the link between the corrugation of the potential and the 
friction force. The ideal way to achieve this goal is provided by a study of 
the load dependence.

We address the load dependence of friction by 
simulating the dynamics of a tip scanning a rigid monolayer graphite surface,
extending to three dimensions ($3D$) 
the Tomlinson-like models of AFM~\cite{tomanek}.
Experimental evidences show that the tip usually cleaves small graphite flakes
which remain attached to it~\cite{dienwiebel}. 
Since the contact diameter between the tip and the surface can be very small
we consider the limiting case in which the tip is formed by a single carbon 
atom connected via harmonic interactions with force constants $K_x$, 
$K_y$ and $K_z$ to a support moving along the scanning direction. 
The carbon atom of coordinates $(x,y,z)$ interacts  
with the graphite surface via the specific 
empirical potential $V_{TS}$ for graphite given by~\cite{los}: 
\begin{equation}
V_{TS}=\sum_{j}\left[\theta(r_0-r_{Tj})V_{1}(r_{Tj})+\theta(r_{Tj}-r_0)V_{2}(r_{Tj})\right] 
\end{equation}
where $\theta(r)$ is the Heaviside function, $r_{Tj}$ is the distance between
the tip carbon atom and the $j-$th substrate atom and 
\begin{equation}
V_i(r)=\epsilon_i\left(e^{-2\beta_i(r-r_0)}-2e^{-\beta_i(r-r_0)}\right)+v_i
\qquad i=1,2
\end{equation}
with $v_1=\epsilon_1-\epsilon_2$ and $v_2=0$.
The values of the parameters are $r_0=0.371$ nm, $\epsilon_1=5.355$ meV, 
$\epsilon_2=2.614$ meV, $\beta_1=14.693$ nm$^{-1}$ and 
$\beta_2=21.029$ nm$^{-1}$.
The tip support is moved with constant velocity $v_{scan}$ along the scanning 
line $x_{scan}=v_{scan}t$, $y_{scan}=$constant.
We solve numerically the equations of motion in the constant force mode, i.e.
we set $K_z=0$ and add a constant force $F_{load}$ in the downward $z$ 
direction, including also a damping term proportional to the atom velocity,
which takes into account dissipation mechanisms:
\begin{equation}
\begin{array}{ccc}
m\ddot{x}=-\frac{\partial V_{TS}}{\partial x}+K_x(x_{scan}-x)-m\eta\dot{x}\\
m\ddot{y}=-\frac{\partial V_{TS}}{\partial y}+K_y(y_{scan}-y)-m\eta\dot{y}\\
m\ddot{z}=-\frac{\partial V_{TS}}{\partial z}-F_{load}-m\eta\dot{z}
\end{array} 
\end{equation}
We adopt an atomistic approach, assuming for $m$ the mass of a single carbon 
atom ($m=1.92\cdot 10^{-26}$ kg) and for the damping parameter 
$\eta=1$ ps$^{-1}$,
which is a value appropriate for dissipation of energy and momentum
at the atomic scale (see e.g.~\cite{ps}).
Here we show results for $K_x=K_y=1$ N/m which are typical values of AFM, 
whereas our scanning velocity $v_{scan}=0.4$ m/s is much higher than in 
experiments. Our choice of parameters makes the dynamics 
underdamped.
\begin{figure}
\epsfig{file=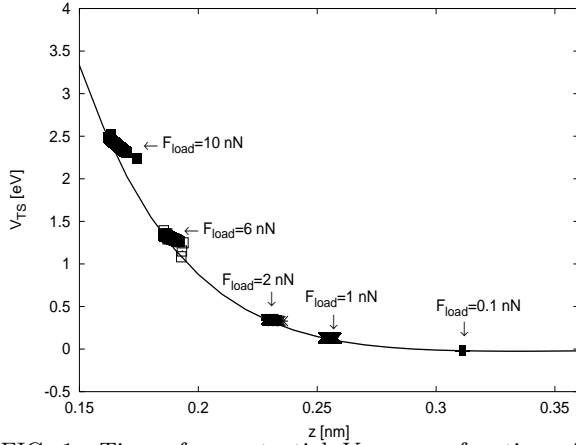,width=8cm,height=6cm}
\caption{Tip-surface potential $V_{TS}$ as a 
function of the tip-surface distance $z$ with $x$ and $y$ coordinates 
taken as those of the hollow site (solid line). 
Points give the values of $V_{TS}$ at several positions along the actual 
trajectory as a function of the corresponding instantaneous value of $z$. 
Results for all indicated loads are obtained for simulations where the 
scanning direction is along a row of atoms ($y_{scan}=0.17$ nm).}
\label{f.potentialzV}
\end{figure}
 
\begin{figure}
\epsfig{file=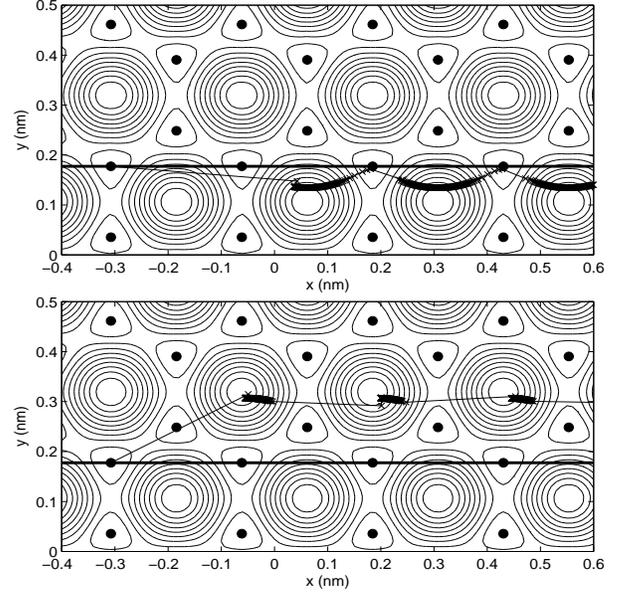,width=8cm,height=8cm}
\caption{Contour plot of the potential $V_{TS}$ for $F_{load}=1$ nN (top) and
$F_{load}=4$ nN (bottom) with $z$ fixed at the average value $<z>$ of the 
height obtained by the simulations ($<z>=0.25$ nm for $F_{load}=1$ nN and 
$<z>=0.2$ nm for $F_{load}=4$ nN). The minimum $V_{min}$ and the maximum 
$V_{max}$ of $V_{TS}$ are located respectively at the hollow site and on top 
of one atom. 
We show 10 contour lines between $V_{min}$ and $V_{max}$, separated by an 
energy interval $\Delta$. $V_{min}=123.5$ meV, $V_{max}=158.5$ meV and 
$\Delta=3.2$ meV for $F_{load}=1$ nN, while $V_{min}=774$ meV, 
$V_{max}=1.01$ eV and $\Delta=23.7$ meV for $F_{load}=4$ nN. The horizontal 
thick solid line indicates the scanning direction ($y_{scan}=0.17$ nm), while 
the crosses are points along the actual atomic trajectory taken every $4$ ps.}
\label{f.contxy}
\end{figure}

The energetics underlying the tip motion for different loads is illustrated 
in Figs.~\ref{f.potentialzV} and~\ref{f.contxy}. In Fig.~\ref{f.potentialzV}
we present the tip-surface potential $V_{TS}$ as a function of the vertical 
tip-surface distance $z$, for in-plane coordinates corresponding to the 
position of the hollow site, on which we superimpose the values of $V_{TS}$
at actual positions of the tip during the simulations for different values 
of the load. 
The tip position probes the attractive part of the potential
for low loads ($F_{load}=0.1$ nN) and moves closer and closer to the 
repulsive core of the substrate atoms for increasing loads. 
Clearly the whole tip-surface potential is
probed by the motion of the tip for different loads, as suggested in
Ref.~\cite{fujisawa2}. 
In Fig.~\ref{f.contxy} we show the contour plots of $V_{TS}$ , calculated for
$z$ given by the average value of the tip height at two values of the load.
The scanning line $y_{scan}=0.17$ nm corresponds to scanning  
along a row of atoms. The actual trajectory is also reported: for the smaller 
load the motion follows a zig-zag-like pattern, while the trajectory 
acquires a one-dimensional stick-slip character for the higher load 
(see also~\cite{fujisawa2}). 

\begin{figure}
\epsfig{file=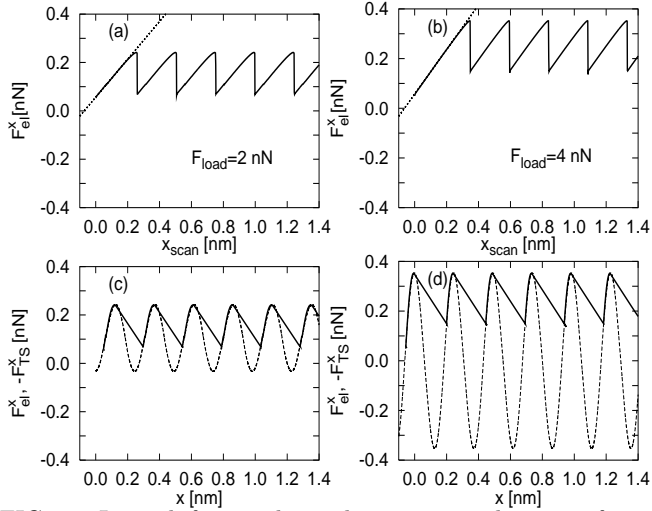,width=9cm,height=7cm}
\caption{Lateral forces along the scanning direction for $F_{load}=2$ nN 
[(a),(c)] and $F_{load}=4$ nN [(b),(d)], plotted as a function of $x_{scan}$
[(a),(b)] and $x$ [(c),(d)], for $y_{scan}=0.17$ nm.
Solid lines are the elastic forces $F_{el}^x$ obtained
by simulations, while the dashed lines in (c) and (d) are static 
calculations of the tip-surface force (with reverted sign) $-F_{TS}^x$ 
at $(y,z)$ determined by averaging $y(t)$ and $z(t)$ 
given by the dynamics. The dotted lines in (a), (b) give the 
slope $K_{eff}$ of the 
sticking part ($K_{eff}=0.78$ N/m for $F_{load}=2$ nN and $K_{eff}=0.89$ N/m 
for $F_{load}=4$ nN).}
\label{f.T0eta1Kxy1v0.0004y0.17F2-4m0.0192}
\end{figure}

Fig.~\ref{f.T0eta1Kxy1v0.0004y0.17F2-4m0.0192} presents 
typical force plots as a function of $x_{scan}$ and $x$~\cite{note} for 
$F_{load}=2$ nN and $F_{load}=4$ nN. The sawtooth behavior
characteristic of stick-slip motion is determined by the competition between 
the elastic force ${\bf F}_{el}$ and the force ${\bf F}_{TS}$ due to the 
tip-surface potential $V_{TS}$. Elastic energy, accumulated 
in the spring, is counterbalanced by the substrate attraction, 
until, suddenly, the tip slips to another minimum.
Therefore, while sticking, ${\bf F}_{el}$ mirrors ${\bf F}_{TS}$. 
This fact can be used to derive $V_{TS}$ itself.
The solid lines in 
Fig.~\ref{f.T0eta1Kxy1v0.0004y0.17F2-4m0.0192} represent 
the lateral force along the scanning direction, $F_{el}^x$, as obtained 
by our simulations. 
Increasing the load enhances the sawtooth behavior of the stick-slip motion
and gives rise to a larger initial sticking region, often observed 
experimentally (see e.g.~\cite{morita}). 

As shown in Fig.~\ref{f.contxy} the actual trajectory does not necessarily
follow  the scanning line so that the potential energy landscape during the 
motion is not known a priori. However, we can extract the effective value of 
the energetic barrier $V_0$ for a given $F_{load}$ directly from the force 
plots. By assuming a sinusoidal shape of $F_{TS}$~\cite{zhong,ke} and noting 
that $F_{TS}$ should average to zero for a periodic substrate,
we can reconstruct $F_{TS}^x$. 
Then, the tip-surface potential $V_{TS}$ is simply given, 
up to a constant, by integrating $F_{TS}^x$: 
\begin{equation}
\label{e.potential}
V_{TS}(x)=-\int F_{TS}^xdx=\frac{V_0}{2}\sin\left(\frac{2\pi x}{a}\right)+const.
\end{equation}
with $V_0=F_0a/(2\pi)$, where $a$ and $F_0$ are the period and the amplitude 
of $F_{TS}$.
In Figs.~\ref{f.T0eta1Kxy1v0.0004y0.17F2-4m0.0192}(c)-(d) we show 
$-F_{TS}^x$ (dashed lines) obtained by static calculations for 
$(y,z)$ determined by averaging $y(t)$ and $z(t)$ given by the dynamics. 
Indeed, $-F_{TS}^x$ follows the sticking parts of $F_{el}^x$ quite well.
The first stick signal, of larger amplitude, is the most suitable
to estimate the amplitude of $F_{TS}^x$. 
\begin{figure}
\epsfig{file=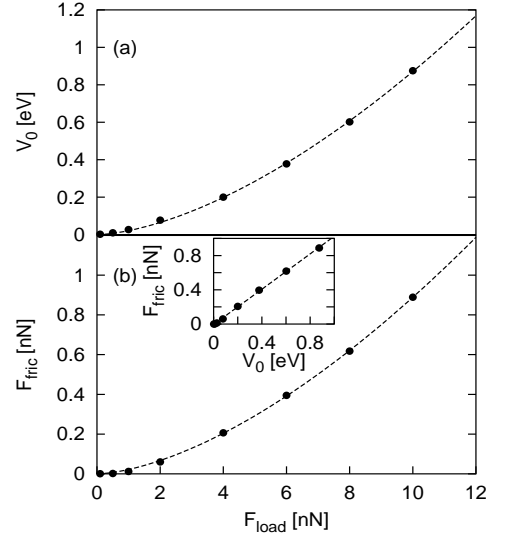,width=12cm,height=8cm}
\caption{Load dependence of the energy barrier (a) and of the friction
force (b) for $y_{scan}=0.17$ nm. Inset of (b): relation between friction
force and energy barrier. The solid circles are the result of the simulations
and the dashed lines are power law fits, as explained in the text.}
\label{f.loadbarr}
\end{figure}

The resulting dependence of the energy barrier $V_{0}$ on the load is shown 
in Fig.~\ref{f.loadbarr}(a). An increase of $V_0$ with $F_{load}$ 
has also been found experimentally~\cite{fujisawa2,riedo}. 
The dashed line in Fig.~\ref{f.loadbarr}(a) is a  power law fit to the 
numerical data with exponent $\sim 1.6$.
Fig.~\ref{f.loadbarr}(b) illustrates the lateral friction force $F_{fric}$, 
obtained as the mean value of the instantaneous lateral force $F_{el}^x$, 
as a function of $F_{load}$. Also these data can be fitted by a 
power law with exponent $\sim 1.6$. 
Thus, the linear relation between macroscopic friction and 
load does not hold at the microscopic level. Moreover, the exponent is 
different from the $2/3$ expected for a Hertzian 
contact~\cite{schwarz,enachescu}. As pointed out in 
Ref.~\cite{schwarz} even a small deviation from the spherical shape of the 
tip can be responsible for a change in the power law exponent. 
We conjecture that exponents larger than one are characteristic of sharp,  
undeformable tip-surface contacts. 
Note that, since the power law
exponents for $V_0$ and $F_{fric}$ as a function of $F_{load}$ are very close,
an approximately linear relation between $F_{fric}$ and $V_0$ is 
recovered (see inset of Fig.~\ref{f.loadbarr}(b)), 
indicating a direct link between atomistic friction and energy barriers for
diffusion. 

In conclusion, we have presented a theoretical study of the $3D$ dynamics of
a tip scanning a graphite surface with realistic 
interactions as a function of the applied load.
We predict a power law behavior with exponent $\sim 1.6$ for the friction 
force as a function of applied load, at variance with macroscopic behavior.
 The study of the load dependence establishes a
direct linear relation between friction and potential corrugation in the 
contact region. 

This work was supported by the Stichting Fundamenteel Onderzoek der Materie
(FOM) with financial support from the Nederlandse Organisatie voor 
Wetenschappelijk Onderzoek (NWO). The authors wish to thank Elisa Riedo,
Jan Los, Sylvia Speller, Sergey Krylov and Jan Gerritsen for useful and 
stimulating discussions.

\end{multicols}


\begin{thebibliography}{99}
\bibitem{note}
Experimental force plots data are often given as a function of $x_{scan}$.
However, since $F_{el}^x=K_x(x_{scan}-x)$, it is straightforward
to rewrite the data $F_{el}^x(x_{scan})$ as $F_{el}^x(x)$.
\end{thebibliography}
\end{document}